\documentclass{czjphys}         
\begin{document}
\title{On (non-Hermitian) Lagrangeans in (particle) physics and
their dynamical generation}  
%
\authori{Frieder Kleefeld\,\footnote{E-mail: {\sf kleefeld@cfif.ist.utl.pt}\,, URL: {\sf http:/$\!$/cfif.ist.utl.pt/$\sim$kleefeld/}}}      \addressi{Centro de
F\'{\i}sica das Interac\c{c}\~{o}es Fundamentais (CFIF), Instituto Superior T\'{e}cnico,\\ Av.\ Rovisco Pais, 1049-001 Lisboa, Portugal}
\authorii{}     \addressii{}
\authoriii{}    \addressiii{}
\authoriv{}     \addressiv{}
\authorv{}      \addressv{}
\authorvi{}     \addressvi{}
%
\headauthor{Frieder Kleefeld}            
\headtitle{On (non-Hermitian) Lagrangeans in (particle) physics
and
their dynamical generation}             
\lastevenhead{Frieder Kleefeld: On (non-Hermitian) Lagrangeans in $\ldots$ physics and their dynamical generation} 
\refnum{A}
\daterec{XXX}    
\issuenumber{0}  \year{2005}
\setcounter{page}{1}
\maketitle

\begin{abstract}
On the basis of a new method to derive the effective action the
nonperturbative concept of ``dynamical generation'' is explained.
A non-trivial, non-Hermitian and PT-symmetric solution for
Wightman's scalar field theory in four dimensions is dynamically
generated, rehabilitating Symanzik's precarious $\phi^4$-theory
with a negative quartic coupling constant as a candidate for an
asymptotically free theory of strong interactions. Finally it is
shown making use of dynamically generation that a Symanzik-like
field theory with scalar confinement for the theory of strong
interactions can be even suggested by experiment.
\end{abstract}

\section{Dynamical generation of Lagrangeans}
\subsection{The concept of dynamical generation}
The concept and terminology of ``dynamical generation'' occurred to us for the first time explicitly in the context of the (one-loop) ``dynamical generation'' of the Quark-Level Linear Sigma Model by M.D.\ Scadron and R.\ Delbourgo \cite{Delbourgo:1993dk}.

A particularly important issue in the process of quantizing a
theory given by some classical Lagrangean is the aspect of
renormalization and renormalizability \cite{Collins:1984xc}. The
process of renormalization is typically performed --- after
choosing some valid regularization scheme (See e.g.\ Ref.\
\cite{Sampaio:2002ii}) --- by adding to the classical Lagrangean
divergent counterterms, which subtract divergencies which would
otherwise show up in the unrenormalized effective action. Naively
one might think that renormalization affects only terms belonging
to the same order of  perturbation theory in some coupling
constant, while other parameters of the same Lagrangean do not
interfere. The underlying philosophy would here be that in a
quantum theory distinct parameters (e.g.\ masses, couplings) in a
Lagrangean can be considered --- like in a classical Lagrangean
--- to a great extent uncorrelated, as long as the Lagrangean is
renormalizable. It appears that this philosophy seems to work
quite well, when it is to renormalize logarithmic divergencies.
That the situation is not so easy can be seen from the formalism
needed to renormalize non-Abelian vector fields
\cite{Veltman:1968ki}. In such theories the values of the coupling
constants responsible for the self-interaction of three vector
fields and of four vector fields are highly correlated due to the
need to cancel appearing quadratic divergencies in the process of
summing up diagrams of different loop order (in particular to
achieve here the fundamental principle of gauge invariance). If
this were not like that, their values could be chosen
independently and therefore also renormalized independently. We
see here a first example of ``dynamical'' generation or
interrelation of two otherwise independent parameters in a
Lagrangean due to the requirement of renormalizability, which
affects here also the {\em cancellation of quadratic
divergencies}. Furthermore we learn that ``dynamical generation''
typically interrelates {\em seemingly uncorrelated parameters of
the Lagrangean} and {\em different loop orders} \footnote{Most
probably the most outstanding example for dynamically generated
theories are theories containing supersymmetry. This is reflected
by the fact that supersymmetric theories typically contain a
minimum of parameters, quadratic divergencies cancel exactly
without extra renormalization (See e.g.\ Ref.\
\cite{Deshpande:1984ke}), and the renormalization of logarithmic
divergencies at one-loop order yields simultaneously an automatic
renormalization of all higher-loop orders. That observation led
already to (non-conclusive) speculations about the question,
whether all theories cancelling quadratic divergencies must be
supersymmetric (See e.g.\ Refs.\ \cite{Deshpande:1984ke}). In
certain situations some --- not necessarily supersymmetric ---
theories may display even strong cancellations on the level of
logarithmic divergencies. In such ``bootstrapping'' theories
physics is determined already at ``tree-level'', as cancelling
loop-contributions show up to be marginal.}. Renormalizable
theories with scalar fields only seem naively to have the
priviledge, not to be affected by the problem faced by non-Abelian
gauge theories, as the quadratic divergencies seem to be
subtractable before entering the renormalization of logarithmic
divergencies. Hence it seems naively, that --- as long as a
Lagrangean with scalar fields only is in a classical sense
considered to be renormalizable --- different parameters of the
Lagrangean can be renormalized individually (up to constraints
resulting from multiplicative renormalization). It is exactly this
misbelief, which leads indeed to the triviality of scalar field
theories like the text book $\phi^4$ theory or even to intimately
related Abelian gauge theories like QED, if not ``dynamically
generated''. If instead the respective theories are ``dynamically
generated'' one {\em does} find --- besides the trivial solution
--- also non-trivial choices of the their parameter space, which
survive the renormalization process without running into
triviality. Interestingly in many cases such non-trivial solutions
are found in the sector of the parameter space related to a {\em
PT-symmetric} \cite{Bender:1998ke}, yet not necessarily to a {\em
Hermitian} non-trivial theory \footnote{Before proceeding we want
to deliver here also some warning about some common regularization
schemes used which must not to be used in the context of
``dynamical generation'': Most important information about
divergencies underlying a theory is contained in tadpole diagrams;
hence any kind of artificial normal ordering or suppression of
important surface terms will erase information needed to dynamical
generate the theory and will lead therefore to wrong conclusions
(See e.g.\ the discussion in Refs.\
\cite{Yang:2001yw,Yang:2002ur}). As dimensional regularization
erases or changes several important divergent diagrams like the
massless tadpole (See e.g.\ Ref.\ \cite{Delbourgo:1998ji}) or the
quadratic divergence in the sunset/sunrise graph (See e.g.\ the
dimensional regularization calculations performed in Refs.\
\cite{Ford:1991hw,Caffo:1998du,Hellman:1986ya}, or on p.\ 114 ff
in Ref.\ \cite{Kleinert:2001ax}), it should not be used to
dynamically generate a theory! According to our experience cutoff
regularization --- if correctly used --- seems to yield always
correct and most compact results compared to other regularization
schemes.}. In order to ``dynamically generate'' a theory (e.g.
like the supersymmetric Wess-Zumino model \cite{Wess:1973kz}) on
the basis of some tentative classical Lagrangean we have to
perform two steps: first we have to construct the terms in the
effective action which are causing non-logarithmic divergencies
(i.e. linear, quadratic, and higher divergencies) in all connected
Feynman-diagrams, which can be constructed from the theory; then
we have to relate and choose the parameters entering these terms
of the effective action such, that all non-logarithmic
divergencies cancel.\footnote{One feels the need to remark that
the very existence of a dynamically generated theory is not always
guaranteed, as the procedure of dynamical generation is intimately
related to renormalization and --- even more --- is strongly
constraining the parameters of the effective action.}

\subsection{New method for the derivation of the effective action and its Lagrangean}
A powerful method to construct the effective action has been known
at least since the benchmarking work of S.\ Coleman \& E.\
Weinberg \cite{Coleman:1973jx} and R.~Jackiw \cite{Jackiw:1974cv}.
Unfortunately it is for our purposes not very convenient, as the
determination of desired terms of the effective action responsible
for leading singularities requires typically the simultaneous
tedious evaluation of many other terms, which do not alter the
discussion.  This is why we want to propose here a different ---
to our best knowledge --- new and more pragmatic approach yielding
equivalent results compared to the formalism of S.\ Coleman, E.\
Weinberg, and R.\ Jackiw. Without loss of generality we want to
explain our simple method here on the basis of some example, the
generalization of which is quite straight forward.

Let's start with the interaction part ${\cal S}_{int}=\int d^4z \;{\cal L}_{int}(\vec{\phi}(z),\partial_z \vec{\phi}(z))$ of an action ${\cal S}$ of $N$ interacting Klein-Gordon fields $\phi_1(z)$, $\ldots$, $\phi_N(z)$. Then the interaction part of the effective action responsible for a process involving $n$ external legs is calculated by the connected ($\left<\ldots\right>_c$) time-ordered vacuum expectation value of the Dyson-operator, where contractions are to be performed over all fields except $n$ fields (``except $\phi^n$''), which remain to be contracted with creation or annihilation operators appearing in initial or final states, i.e.:
\begin{eqnarray} \frac{i}{1!} \; {\cal S}_{e\!f\!f} & = & \left<0\right| T [\,\exp(i\,{\cal S}_{int})] \left|0\right>_c|_{\mbox{\small except}\; \phi^n} \nonumber \\
 & = & \frac{i}{1!} \; \left<0\right| T [\,{\cal S}_{int}] \left|0\right>_c|_{\mbox{\small except}\; \phi^n} + \frac{i^2}{2!} \; \left<0\right| T [\,{\cal S}_{int}\, {\cal S}_{int}] \left|0\right>_c|_{\mbox{\small except}\; \phi^n} + \nonumber \\
 & + & \frac{i^3}{3!} \; \left<0\right| T [\,{\cal S}_{int}\, {\cal S}_{int}\, {\cal S}_{int}] \left|0\right>_c|_{\mbox{\small except}\; \phi^n} + \ldots \; .
\end{eqnarray}
The method is proved by making heavy use of the following identity (inserted between initial and final states $\left|i\right>$ and $\left<f\right|$, respectively) found e.g. on p.\ 44 in a well known book by C.\ Nash \cite{Nash:1978bg}, i.e.:
\begin{eqnarray} \lefteqn{\left<f\right| T [\,\exp(i\,{\cal S}_{int})] 
\left|i\right> \; = \; \left<f\right| \exp\Bigg(\left[ \frac{1}{2} \, 
\left<\phi^2\right> \frac{\delta^2}{\delta\phi^2}\right] \Bigg) \; : 
\exp(i\,{\cal S}_{int}) : \left|i\right> \; =} \nonumber \\
  & = & \left<f\right| \exp\Bigg(\left[ \frac{1}{2} \left<\phi^2\right> 
\frac{\delta^2}{\delta\phi^2}\right] \Bigg) \; \Big( : \frac{i}{1!}\,{\cal 
S}_{int} : + : \frac{i^2}{2!}\,{\cal S}_{int}{\cal S}_{int} : + \ldots 
\Big) \left|i\right> \nonumber \\
  & = & \left<f\right| \Big[ \frac{i}{1!} \Big\{ : {\cal S}_{int} : + 
\frac{1}{1!} \left[ \frac{1}{2} \left<\phi^2\right> 
\frac{\delta^2}{\delta\phi^2}\right] : {\cal S}_{int} : + \frac{1}{2!} 
\left[ \frac{1}{2} \left<\phi^2\right> 
\frac{\delta^2}{\delta\phi^2}\right]^2 : {\cal S}_{int} : + \ldots \Big\} 
\nonumber \\
  & & \makebox[2mm]{} + \frac{i^2}{2!} \Big\{ : {\cal S}^2_{int} : + 
\frac{1}{1!} \left[ \frac{1}{2} \left<\phi^2\right> 
\frac{\delta^2}{\delta\phi^2}\right] : {\cal S}^2_{int} : + \frac{1}{2!} 
\left[ \frac{1}{2} \left<\phi^2\right> 
\frac{\delta^2}{\delta\phi^2}\right]^2 : {\cal S}^2_{int} : + \ldots 
\Big\} \nonumber \\
  & & \makebox[2mm]{} + \frac{i^3}{3!} \Big\{ : {\cal S}^3_{int} : + 
\frac{1}{1!} \left[ \frac{1}{2} \left<\phi^2\right> 
\frac{\delta^2}{\delta\phi^2}\right] : {\cal S}^3_{int} : + \frac{1}{2!} 
\left[ \frac{1}{2} \left<\phi^2\right> 
\frac{\delta^2}{\delta\phi^2}\right]^2 : {\cal S}^3_{int} : + \ldots 
\Big\} \nonumber \\
  & & \makebox[2mm]{} + \ldots \Big] \left|i\right> \; ,
\end{eqnarray}
where we have defined for convenience the short-hand notation
\begin{equation} \left[ \frac{1}{2} \left<\phi^2\right> \frac{\delta^2}{\delta\phi^2}\right] \equiv \frac{1}{2} \, \sum\limits^N_{i_1,i_2=1}\int d^4z_1 d^4z_2 \left<0\right| T [\,\phi_{i_1}(z_1) \, \phi_{i_2}(z_2)] \left|0\right> \frac{\delta^2}{\delta \phi_{i_2}(z_2) \delta \phi_{i_1}(z_1)} \; .
\end{equation}
The identity (See e.g.\ p.\ 49 in Ref.\ \cite{Nash:1978bg}) and
method is easily extended to Fermions, i.e. Grassmann fields
$\psi_{1}(z)$, \ldots, $\psi_{N}(z)$, by replacing $\left[
\frac{1}{2} \left<\phi^2\right>
\frac{\delta^2}{\delta\phi^2}\right]$ by
\begin{equation} \left[ \left<\psi\,\bar{\psi}\right> \frac{\delta^2}{\delta\bar{\psi}\delta\psi}\right] \equiv \sum\limits^N_{i_1,i_2=1}\int d^4z_1 d^4z_2 \left<0\right| T [\psi_{i_1}(z_1)  \bar{\psi}_{i_2}(z_2)] \left|0\right> \frac{\delta^2}{\delta \bar{\psi}_{i_2}(z_2) \delta \psi_{i_1}(z_1)} \; .
\end{equation}
Convince yourself, that the method reproduces S.\ Coleman's and E.\ Weinberg's loop-expansion \cite{Coleman:1973jx} for a simple massless $\phi^4$-theory with ${\cal S}_{int}=\int d^4z \; (-\frac{\lambda}{4!})\,\phi^4(z)$~.\footnote{We show here only the most important steps of the derivation:
\begin{eqnarray} \frac{i}{1!} \; {\cal S}_{e\!f\!f} & = & 
\sum\limits_{n=1}^\infty \frac{i^n}{n!} \left<0\right| T [\,{\cal 
S}^n_{int}] \left|0\right>_c|_{\mbox{\small except}\; \phi^{2n}}  = 
\sum\limits_{n=1}^\infty \frac{i^n}{n!} \, \frac{1}{n!}\left[ \frac{1}{2} 
\left<\phi^2\right> \frac{\delta^2}{\delta\phi^2}\right]^n {\cal 
S}^n_{int}  \nonumber \\
  & = & \sum\limits_{n=1}^\infty \frac{i^n}{n!} \, \frac{1}{n!} \; \int 
d^4z_1 \ldots d^4z_n \; \frac{n! (n-1)!}{2} \; 
\left(-\frac{\lambda}{2!}\right)^n\,\phi^2(z_1) \ldots \phi^2(z_n) \times 
\nonumber \\
  & & \times \left<0\right| T [\,\phi(z_1) \, \phi(z_2)] 
\left|0\right>\left<0\right| T [\,\phi(z_2) \, \phi(z_3)] \left|0\right> 
\ldots \left<0\right| T [\,\phi(z_n) \, \phi(z_1)] \left|0\right> \nonumber 
\\
  & = & \sum\limits_{n=1}^\infty 
\left(\frac{\lambda}{2!}\right)^n\,\frac{1}{2 n} \, \int d^4z_1 \ldots 
d^4z_n \;\phi^2(z_1) \ldots \phi^2(z_n) \times \nonumber \\
  & & \times \int 
\frac{d^4p_{12}}{(2\pi)^4}\frac{d^4p_{23}}{(2\pi)^4}\ldots 
\frac{d^4p_{n1}}{(2\pi)^4}  \frac{e^{-ip_{12}(z_1-z_2)} 
e^{-ip_{23}(z_2-z_3)} \ldots e^{-ip_{n1}(z_n-z_1)}}{(p^2_{12} 
+i\varepsilon)(p^2_{23} +i\varepsilon)\ldots (p^2_{n1} +i\varepsilon)} 
\nonumber \\
  & = & \int d^4z \left( \sum\limits_{n=1}^\infty 
\left(\frac{\lambda}{2!}\right)^n\frac{1}{2 n}  \int \frac{d^4p}{(2\pi)^4} 
\left(\frac{\phi^2(0)}{p^2 +i\varepsilon}\right)^n + \mbox{non-local 
terms} \right) .
\end{eqnarray}
Some of the resulting non-local terms are nicely discussed e.g.\ in Ref.\ \cite{Cheyette:1985ue}.}
\section{Applications}
\subsection{A.S.\ Wightman's (non-)trivial and K.\ Symanzik's precarious $\phi^4$ theory}
In this section we want to shortly sketch the steps to dynamically
generate the ``Scalar Wightman Theory in 4 Space-Time Dimensions''
\cite{Streater:2005a} (See also Ref.~\cite{Kleinert:2001ax}). As
we will see below, the dynamical generation of this so-called
$\phi^4$ theory yields --- besides the well known ``trivial''
solution --- the ``precarious'' \cite{Stevenson:1985zy}
non-trivial solution suggested by K.\ Symanzik
\cite{Symanzik:1971a} being non-Hermitian and --- under certain
circumstances also --- PT-symmetric \cite{Bender:1998ke}.

To dynamically generate a $\phi^N$-theory upto $N=4$ we start from the following lowest order action containing just a three-point interaction:
\begin{eqnarray} {\cal S}_{{}_{(0)}} & = & \int d^4z \; \left\{\, \frac{1}{2} \Big( (\partial\, \phi_{{}_{(0)}}(z))^2 - m^2_{{}_{(0)}}\;\phi^2_{{}_{(0)}}(z) \Big) \; - \, \frac{1}{3!} \; g_{{}_{(0)}} \; \phi^3_{{}_{(0)}}(z) \, \right\} \nonumber \\[1mm]
 & = & {\cal S}_{{}_{(0)}}[\,(\partial\, \phi)^2\,] + {\cal S}_{{}_{(0)}}[\,\phi^2\,] + {\cal S}_{{}_{(0)}}[\,\phi^3\,] \; .
\end{eqnarray}
In a first step we want to absorb by dynamical generation the finite one-loop correction to the $\phi^3$-coupling into a renormalization of the three-point coupling, i.e.:
\begin{eqnarray} \frac{i}{1!}\;  {\cal S}_{{}_{(1)}}[\,\phi^3\,]  & = & \frac{i}{1!} \, \left<0\right|\! T \Big[{\cal S}_{{}_{(0)}}[\,\phi^3\,]\,\Big]\left|0\right>_c\Big|_{\mbox{except $\,\phi^3_{{}_{(0)}}$}} \nonumber \\
 & + & \frac{i^3}{3!} \, \left<0\right|\! T \Big[{\cal S}_{{}_{(0)}}[\,\phi^3\,]\;{\cal S}_{{}_{(0)}}[\,\phi^3\,]\;{\cal S}_{{}_{(0)}}[\,\phi^3\,]\,\Big]\left|0\right>_c\Big|_{\mbox{except $\,\phi^3_{{}_{(0)}}$}} \; .
\end{eqnarray}
The next step is to dynamically generate on the basis of ${\cal S}_{{}_{(1)}}[\,\phi^3\,]$ the term of the effective action quadratic in the fields $\phi_{{}_{(0)}}(z)$ {\em assuming} the absence of quadratically divergent terms. \footnote{I.e.\ we consider:
\begin{eqnarray} \lefteqn{\frac{i}{1!}\; \Big( {\cal S}_{{}_{(1)}}[\,(\partial \phi)^2\,] + {\cal S}_{{}_{(1)}}[\,\phi^2\,] \, \Big) = \frac{i}{1!} \, \left<0\right|\! T \Big[{\cal S}_{{}_{(0)}}[\,(\partial \phi)^2\,] \,\Big]\left|0\right>_c\Big|_{\mbox{except $\,\phi^2_{{}_{(0)}}$}} +} \nonumber \\
 & + & \frac{i}{1!} \, \left<0\right|\! T \Big[{\cal S}_{{}_{(0)}}[\,\phi^2\,] \,\Big]\left|0\right>_c\Big|_{\mbox{except $\,\phi^2_{{}_{(0)}}$}}
 + \frac{i^2}{2!} \, \left<0\right|\! T \Big[{\cal S}_{{}_{(1)}}[\,\phi^3\,]\,{\cal S}_{{}_{(1)}}[\,\phi^3\,]\,\Big]\left|0\right>_c\Big|_{\mbox{except $\,\phi^2_{{}_{(0)}}$}} \!\!.
\end{eqnarray}}
The result of the previous steps is simple multiplicative
coupling, wave function and mass renormalization, as we obtain as
a whole (The omissions (``$\ldots$'') denote here non-local terms
not relevant for our present discussion.):
\begin{eqnarray} \lefteqn{{\cal S}_{{}_{(1)}}[\,(\partial \phi)^2\,] + {\cal S}_{{}_{(1)}}[\,\phi^2\,] + {\cal S}_{{}_{(1)}}[\,\phi^3\,]\; =} \nonumber \\
 & = & \int d^4z \left( \frac{1}{2} \Big( (\partial\, \phi_{{}_{(1)}}(z))^2 - m^2_{{}_{(1)}}\,\phi^2_{{}_{(1)}}(z) \Big) - \frac{1}{3!} \, g_{{}_{(1)}} \, \phi^3_{{}_{(1)}}(z) \right) + \ldots\; ,
\end{eqnarray}
with
\begin{eqnarray}
g_{{}_{(1)}} & = & \bar{g}_{{}_{(0)}} \Bigg/\left( 1 - \, \frac{1}{32\,\pi^2} \;  \frac{\bar{g}^2_{{}_{(0)}}}{m^2_{{}_{(0)}}} \right)^{3/2} \; , \;
\bar{g}_{{}_{(0)}} \; = \; g_{{}_{(0)}}  \, \left( 1+  \frac{1}{32\,\pi^2} \; \frac{g^2_{{}_{(0)}}}{m^2_{{}_{(0)}}} \right) \; ,  \nonumber \\
\phi^2_{{}_{(1)}}(z) & = & \phi^2_{{}_{(0)}}(z) \; \left( 1 - \,\frac{1}{32\,\pi^2} \;  \frac{\bar{g}^2_{{}_{(0)}}}{m^2_{{}_{(0)}}} \right) \; ,\nonumber \\
 m^2_{{}_{(1)}} & = & m^2_{{}_{(0)}}\; \left( 1 + \frac{i}{2} \;  \frac{\bar{g}^2_{{}_{(0)}}}{m^2_{{}_{(0)}}} \int \frac{d^4p}{(2\pi)^4}\;\frac{1}{(p^2- m^2_{{}_{(0)}})^2} \right)\Bigg/\left( 1 - \, \frac{1}{32\,\pi^2} \;  \frac{\bar{g}^2_{{}_{(0)}}}{m^2_{{}_{(0)}}} \right) \; . \nonumber \\ \label{repleqns1}
\end{eqnarray}
If we renormalize this result through a suitable mass counter term yielding a log.-divergent gap-equation promoted e.g.\ by M.D.\ Scadron \cite{Hakioglu:1990kg}, i.e.\ by applying
\begin{equation} \int \frac{d^4p}{(2\pi)^4}\;\frac{1}{(p^2- m^2_{{}_{(0)}})^2} \; \longrightarrow \; +\, \frac{i}{16\,\pi^2} \; , \label{logdiveq1}
\end{equation}
then we have a bootstrapping situation for the mass, as there holds then $m^2_{{}_{(1)}} =  m^2_{{}_{(0)}}$. Recall that the result has been obtained by {\em assuming} the absence, i.e. the {\em cancellation} of quadratically divergent terms in ${\cal S}_{{}_{(1)}}[\,(\partial \phi)^2\,] + {\cal S}_{{}_{(1)}}[\,\phi^2\,]$.
In order to show now the absence of quadratically divergent terms for self-consistency reasons, we have first to dynamically generate on the basis of $g_{{}_{(1)}}$ and $m_{{}_{(1)}}$ the effective action for a four-point interaction of the field $\phi_{{}_{(0)}}(z)$, and then test the cancellations of quadratic divergencies on the level of tadpoles and selfenergies. The effective action for a four-point interaction of the field $\phi_{{}_{(0)}}(z)$ (expressed in terms of $\phi_{{}_{(1)}}(z)$) is here dynamically generated  {\em for simplicity just} up to order $g^4_{{}_{(1)}}$ {\em assuming} again the absence of quadratically divergent terms, i.e.:
\begin{eqnarray} \lefteqn{\frac{i}{1!}\;  {\cal S}_{{}_{(1)}}[\,\phi^4\,] 
\; = \;\frac{i}{1!}\;   \left( {\cal S}^{tree\,}_{{}_{(1)}}[\,\phi^4\,] + 
{\cal S}^{loop\,}_{{}_{(1)}}[\,\phi^4\,]\right) \; =} \nonumber \\
  & = &  \frac{i^2}{2!} \, \left<0\right|\! T \Big[{\cal 
S}_{{}_{(1)}}[\,\phi^3\,]\;{\cal 
S}_{{}_{(1)}}[\,\phi^3\,]\,\Big]\left|0\right>_c\Big|_{\mbox{except 
$\,\phi^4_{{}_{(1)}}$}} \nonumber \\
  & + &  \frac{i^4}{4!} \, \left<0\right|\! T \Big[{\cal 
S}_{{}_{(1)}}[\,\phi^3\,]\;{\cal S}_{{}_{(1)}}[\,\phi^3\,]\;{\cal 
S}_{{}_{(1)}}[\,\phi^3\,]\;{\cal 
S}_{{}_{(1)}}[\,\phi^3\,]\,\Big]\left|0\right>_c\Big|_{\mbox{except 
$\,\phi^4_{{}_{(1)}}$}} \nonumber \\
  & = & \frac{i^2}{2!} \, \int d^4z_1 \, d^4z_2 \;\left( - \frac{1}{3!} \, 
g_{{}_{(1)}} \right)^2  \, 3^2 \, \phi^2_{{}_{(1)}}(z_1)\, 
\phi^2_{{}_{(1)}}(z_2) \; i \int \frac{d^4p}{(2\pi)^4} \,  \frac{e^{-ip 
(z_1-z_2)}}{(p^2- m^2_{{}_{(1)}})} \nonumber \\
  & + & \frac{i^4}{4!} \int d^4z_1 d^4z_2 d^4z_3 d^4z_4 \left( - 
\frac{1}{3!} g_{{}_{(1)}} \right)^4  3 (3!)^4 \phi_{{}_{(1)}}(z_1) 
\phi_{{}_{(1)}}(z_2)\phi_{{}_{(1)}}(z_3) \phi_{{}_{(1)}}(z_4) \, i^4 
\times \nonumber \\
  & \times & \int\! \frac{d^4p_{{}_{12}}}{(2\pi)^4} 
\frac{d^4p_{{}_{23}}}{(2\pi)^4} \frac{d^4p_{{}_{34}}}{(2\pi)^4} 
\frac{d^4p_{{}_{41}}}{(2\pi)^4} \frac{e^{-ip_{{}_{12}} (z_1-z_2)} 
e^{-ip_{{}_{23}} (z_2-z_3)} e^{-ip_{{}_{34}} (z_3-z_4)} e^{-ip_{{}_{41}} 
(z_4-z_1)}}{(p^2_{{}_{12}}- m^2_{{}_{(1)}})(p^2_{{}_{23}}- 
m^2_{{}_{(1)}})(p^2_{{}_{34}}- m^2_{{}_{(1)}})(p^2_{{}_{41}}- 
m^2_{{}_{(1)}})} \nonumber \\
  & = & \frac{i}{1!} \, \int d^4z_1\, d^4z_2  \left( - \frac{1}{4!}\right) 
\, 3\, g^2_{{}_{(1)}}  \; \phi^2_{{}_{(1)}}(z_1)\, \phi^2_{{}_{(1)}}(z_2) 
\, \int \frac{d^4p}{(2\pi)^4}\;  \frac{e^{-ip (z_1-z_2)}}{(p^2- 
m^2_{{}_{(1)}})} \nonumber \\
  & + & \frac{i}{1!} \,  \int d^4z \, \left( - \frac{1}{4!} \right) \, 3\, 
i \, g^4_{{}_{(1)}} \; \phi^4_{{}_{(1)}}(z) \, \int 
\frac{d^4p}{(2\pi)^4}\;  \frac{1}{(p^2- m^2_{{}_{(1)}})^4} + \ldots 
\nonumber \\
  & = & \frac{i}{1!}\int d^4z \, \left( -\, \frac{1}{4!}\right) \left( 
(-\,3)\;\, \frac{g^2_{{}_{(1)}}}{m^2_{{}_{(1)}}} \, \phi^4_{{}_{(1)}}(z) + 
\left( - \frac{1}{32\;\pi^2}\right) \, 
\frac{g^4_{{}_{(1)}}}{m^4_{{}_{(1)}}} \, \phi^4_{{}_{(1)}}(z)  \right)  + 
\ldots \nonumber \\
  & = & \frac{i}{1!} \, \int d^4z \left( \left( -\, \frac{1}{4!}\right) \; 
(-\,3)\;\, \frac{g^2_{{}_{(1)}}}{m^2_{{}_{(1)}}}  \, \phi^4_{{}_{(1)}}(z) 
+ \left( - \, \frac{1}{4!} \, \lambda_{{}_{(1)}} \right) \, 
\phi^4_{{}_{(1)}}(z) \right) + \ldots \; .
\end{eqnarray}
As a result of this consideration we have
\begin{equation} {\cal S}_{{}_{(1)}} = \int d^4z \left( \frac{1}{2} \Big( (\partial \phi_{{}_{(1)}}(z))^2 - m^2_{{}_{(1)}} \phi^2_{{}_{(1)}}(z) \Big) -  \frac{1}{3!} g_{{}_{(1)}} \phi^3_{{}_{(1)}}(z) -  \frac{1}{4!} \lambda_{{}_{(1)}} \phi^4_{{}_{(1)}}(z)\right) + \ldots\, ,
\end{equation}
with $\lambda_{{}_{(1)}} =   -  g^4_{{}_{(1)}}/(32\pi^2\,m^4_{{}_{(1)}})$ and the replacements made in Eq.\ (\ref{repleqns1}).
Let's see now on the basis of this action, in how far quadratic divergencies cancel, as assumed in our approach from the beginning. Therefore we dynamically generate --- for convenience --- e.g.\ the effective action describing the sum of quadratically divergent tadpoles:
\begin{eqnarray} \frac{i}{1!}\;  {\cal S}_{{}_{(1)}}[\,\phi\,]
 & = &  \frac{i}{1!} \, \left<0\right|\! T \Big[{\cal S}_{{}_{(1)}}[\,\phi^3\,]\,\Big]\left|0\right>_c\Big|_{\mbox{except $\,\phi_{{}_{(1)}}$}} \nonumber \\
 & + & \frac{i^2}{2!} \, 2!\, \left<0\right|\! T \Big[{\cal S}^{loop\,}_{{}_{(1)}}[\,\phi^4\,]\;{\cal S}_{{}_{(1)}}[\,\phi^3\,]\,\Big]\left|0\right>_c\Big|_{\mbox{except $\,\phi_{{}_{(1)}}$}} \nonumber \\
 & = & \frac{i}{1!}  \, \int d^4z \;\left( - \, \frac{1}{3!} \; g_{{}_{(1)}} \right)  \; 3 \; \,\phi_{{}_{(1)}}(z) \; \; i \int \frac{d^4p}{(2\pi)^4} \;  \frac{1}{(p^2- m^2_{{}_{(1)}})} \nonumber \\[2mm]
 & + & \frac{i}{1!} \, \, \int d^4z \;\left( - \, \frac{1}{3!} \; g_{{}_{(1)}} \right) \;  (-1) \; \lambda_{{}_{(1)}} \; \phi_{{}_{(1)}}(z) \; \times \nonumber \\
 & \times & \int \frac{d^4p_{{}_{1}}}{(2\pi)^4} \frac{d^4p_{{}_{2}}}{(2\pi)^4} \frac{d^4p_{{}_{3}}}{(2\pi)^4}\;  \frac{(2\pi)^4 \;\delta^4(p_{{}_{1}}+p_{{}_{2}}+p_{{}_{3}})}{(p^2_{{}_{1}}- m^2_{{}_{(1)}})(p^2_{{}_{2}}- m^2_{{}_{(1)}})(p^2_{{}_{3}}- m^2_{{}_{(1)}})} \; . \label{tadpoleeq1}
\end{eqnarray}
To proceed further we extract shortly in the footnote the leading
singularity structure of the occuring massive sunset/sunrise
diagram, being particularly complicated due to the overlap of one
quadratic divergence with three logarithmic divergences (See e.g.\
p.\ 78 ff in Ref.\ \cite{Nash:1978bg}).\footnote{The safest and
most compact discussion of the sunset/sunrise diagram is achieved
in cutoff regularization, even though the full diagram in cutoff
regularization has --- to our present knowledge --- never been
calculated in a closed form. For a discussion of the finite part
of the sunset/sunrise integral for non-zero external four-momentum
on the basis of implicit renormalization see e.g.\ Ref.\
\cite{Sampaio:2002ii}. The leading divergent parts of the
sunset/sunrise diagram for zero external four-momentum and equal
masses have been determined in cutoff regularization in Ref.\
\cite{Yang:2003bv} to be:
\begin{eqnarray} \lefteqn{\int^\Lambda \frac{d^4p_1}{(2\pi)^4} \int^\Lambda \frac{d^4p_2}{(2\pi)^4} \int^\Lambda \frac{d^4p_3}{(2\pi)^4} \, \frac{(2\pi)^4\; \delta^4 (p_1+p_2+p_3)}{(p_1^2- m^2)(p_2^2- m^2)(p_3^2- m^2)}\; =} \nonumber \\
 & = & - \left( \frac{1}{16\,\pi^2} \right)^2  \left( 2 \Lambda^2 + \frac{3}{2} m^2 \ln^2\left( \frac{\Lambda^2}{m^2} \right) -  3 m^2\, \ln\left( \frac{\Lambda^2}{m^2} \right) + C\, m^2 \right)  + O(\Lambda^{-2})\; , \label{sunsetcutoffeq1}
\end{eqnarray}
while the integration constant $C$ was numerically estimated in Ref.\ \cite{Yang:2003bv} to be approximately $C\simeq 4$. After recalling $\int^\Lambda \frac{d^4p}{(2\pi)^4}  \frac{1}{(p^2- m^2)^2} = \frac{i}{16\, \pi^2}  \left( \ln \frac{\Lambda^2+m^2}{m^2}  - \frac{\Lambda^2}{\Lambda^2+m^2} \right)$ and $\int^\Lambda \frac{d^4p}{(2\pi)^4} \frac{1}{(p^2- m^2)} =\frac{-i}{16 \pi^2} \;m^2 \left( \frac{\Lambda^2}{m^2} - \ln \frac{\Lambda^2 + m^2}{m^2} \right)$ Eq.\ (\ref{sunsetcutoffeq1}) is replaced for $\Lambda\rightarrow \infty$ and in the local limit by
\begin{eqnarray} \lefteqn{I_{sunset/sunrise}=\int \frac{d^4p_1}{(2\pi)^4} \frac{d^4p_2}{(2\pi)^4} \frac{d^4p_3}{(2\pi)^4} \, \frac{(2\pi)^4\; \delta^4 (p_1+p_2+p_3)}{(p_1^2- m^2)(p_2^2- m^2)(p_3^2- m^2)} =} \nonumber \\
 & = &  - 2 \; \frac{i}{16\pi^2} \int \frac{d^4p}{(2\pi)^4} \frac{1}{(p^2- m^2)} +  \frac{2}{3} \; m^2 \left( \frac{3}{2} \int \frac{d^4p}{(2\pi)^4}   \frac{1}{(p^2- m^2)^2} +  \frac{i}{16\,\pi^2} \right)^2  \nonumber \\
 & & + \left( \frac{1}{16\pi^2} \right)^2  m^2  \left(\frac{1}{6} - \; C\right)  +\ldots \; .
\end{eqnarray}
The last line displays the most divergent part of the massive sunset/sunrise diagram at zero external four-momentum in a regularization scheme independent manner. The application of a renormalization scheme yielding the ``bootstrapping'' log.-divergent gap-equation Eq.\ (\ref{logdiveq1}) reduces the foregoing equation finally to
\begin{equation}
I_{sunset/sunrise} \rightarrow  - 2\, \frac{i}{16\pi^2} \int \frac{d^4p}{(2\pi)^4} \,  \frac{1}{(p^2- m^2)}  - \left( \frac{1}{16\pi^2} \right)^2 \, m^2 \, (4 +  C)  +\ldots \; .
\end{equation}}
The expression for the leading divergence of the sunset/sunrise graph is then to be inserted in Eq.\ (\ref{tadpoleeq1}) yielding the following result for the local limit of the effective action describing tadpoles:
\begin{eqnarray} \lefteqn{{\cal S}_{{}_{(1)}}[\,\phi\,] = \int d^4z \, \left( - \frac{1}{3!} \, g_{{}_{(1)}} \right)  3\,i \; \phi_{{}_{(1)}}(z) \;\times} \nonumber \\
 & \times & \Bigg\{ \left( 1+\frac{2}{3}\, \frac{1}{16\pi^2} \, \lambda_{{}_{(1)}}\right)  \int \frac{d^4p}{(2\pi)^4} \,  \frac{1}{(p^2- m^2_{{}_{(1)}})}  -i\, \left( \frac{1}{16\pi^2} \right)^2  m^2_{{}_{(1)}}  \frac{(4 +  C)}{3} \,\lambda_{{}_{(1)}} \Bigg\}\nonumber \\
 & + & \ldots \; .
\end{eqnarray}
Simple inspection of this expression yields that the quadratic
divergencies cancel on one hand for the well known ``trivial''
solution $g_{{}_{(1)}}=0$. On the other hand the dynamically
generated theory displays a non-trivial, precarious solution in
the spirit of K.~Symanzik for $\lambda_{{}_{(1)}}=-(3/2)\,
16\pi^2=-24\pi^2$ implying due to $\lambda_{{}_{(1)}} =  -
g^4_{{}_{(1)}}/(32\pi^2 m^4_{{}_{(1)}})$ four solutions for the
three-point coupling constant $g_{{}_{(1)}}$, i.e.
$g_{{}_{(1)}}=\pm 4\pi \, 3^{1/4} \, m_{{}_{(1)}}$ and
$g_{{}_{(1)}}=\pm i\, 4\pi \, 3^{1/4} \, m_{{}_{(1)}}$.
Furthermore we notice that for the probable case of $C\not=-4$ and
non-vanishing mass $m_{{}_{(1)}}$ the non-trivial theory develops
already at this stage a finite non-vanishing vacuum expectation
value (See also the discussion in Ref.\ \cite{Yang:2003bv}).
Finally we mention in view of self-consistency without listing the
explicit proof that the obtained non-trivial values for
$\lambda_{{}_{(1)}}$ and $g_{{}_{(1)}}$ lead also to a
cancellation of quadratic divergencies on the level of the
selfenergy, consistent with our starting assumption that quadratic
divergencies cancel.

\subsection{A non-Hermitian and ``PT-symmetric'' theory of strong interactions}

The purpose of this section is to demonstrate on the basis of
experimental ``evidence'' that a dynamically generated theory of
strong interactions based on mesons and quarks has to be
non-Hermitian and close to PT-symmetric \cite{Bender:1998ke}.
Starting point for our considerations --- inspired somehow by
Ref.\ \cite{Cabibbo:1970uc} --- is the sum of the interaction
Lagrangean of weak interactions containing (anti)leptons denoted
by $\ell_-(x)$, $\overline{\ell^c_+}(x)$ and (anti)quarks denoted
by $q_-(z)$, $\overline{q^c_+} (z)$ and a Yukawa-like interaction
Lagrangean describing the strong interaction between (anti)quarks
and scalar ($S(z)$), pseusoscalar ($P(z)$), vector ($V(z)$), and
axialvector ($Y(z)$) $U(6)\times U(6)$ meson field matrices in
flavour space inspired by Ref.\ \cite{Levy:1967a} (See also
\cite{Kleefeld:2002au,Kleefeld:2004jb}) (The undetermined signs
$s_s$, $s_p$, $s_v$, $s_y\in \{-1,+1\}$ are here irrelevant!):
\begin{eqnarray} \lefteqn{{\cal L}^{\,strong}_{\,int} (z) =} \nonumber \\
 & = &  \sqrt{2}\, g \;\, \overline{q^c_+} (z)  \left( s_s \, S(z) + s_p \,i\,P(z)\gamma_5 + \frac{e^{-i\,\alpha}}{2}  \Big(s_v \not\!{V}(z) + s_y \not\!{Y}(z)\, \gamma_5\Big) \right)  q_-(z) \; , \nonumber \\
\end{eqnarray}
with $g=|g|\exp(i\alpha)$ being the eventually complex strong interaction coupling constant, while contrary to Refs.\ \cite{Cabibbo:1970uc,Levy:1967a} we {\em do not allow any further extra direct meson-meson interaction terms} in the Lagrangean, as they shall be generated dynamically through quark-loops only \footnote{This follows the same philosophy as in the previous section, where the $\phi^4$-interaction was dynamically generated starting out just from a $\phi^3$-theory. It is an interesting possibility to be considered in future, whether in a similar manner the whole non-Fermionic part of the Lagrangean of the standard model of particle physics can be dynamically generated on the basis of Yukawa-like interaction terms coupling of Bosons (gauge-bosons, Higgs-(pseudo)scalars, $\dots$) to Fermions, i.e. (anti)quarks and (anti)leptons.}.
The first step is now to study leptonic decays of pseudoscalar mesons to extract the pseudoscalar decay constants $f_P$. By dynamical generation we obtain for the relevant part of the effective action $S_{e\!f\!f}$ in the local limit ($M_q\equiv\mbox{diag}[m_u,m_c,m_t,m_d,m_s,m_b]$, ``$\mbox{tr}_{{}_F}$''= flavour trace) \footnote{We assumed here without loss of generality for traditional reasons a colour factor $N_c$, which can be absorbed by a redefinition of the strong coupling constant $g$.}:
\begin{eqnarray} \lefteqn{\frac{i}{1!} \, {\cal S}_{e\!f\!f} = \frac{i}{1!} \left<0\right|\! T [S\,]\left|0\right>_c\Big|_{\mbox{except $P\bar{\ell}\ell\,$}} + \frac{i^2}{2!} \left<0\right|\! T [{\cal S}\,{\cal S}\,]\left|0\right>_c\Big|_{\mbox{except $P\,\bar{\ell}\ell\,$}} + \ldots} \nonumber \\
 & = &  \int d^4z \; \left( -\, 2\, \frac{G_F}{\sqrt{2}}\right) \, \sqrt{2} \;s_p \;\, e^{i\,\alpha} \times \nonumber \\[1mm]
 & \times & \mbox{tr}_{{}_F}\Big[ \, -\, 4\,i\, N_c \, |g|\, \int \frac{d^4p}{(2\pi)^4} \;  \frac{1}{p^2- M_q^2 } \,\;  \frac{1}{2} \; \{M_q\,, \, (\partial_\mu P(z)) \} \; \frac{1}{p^2- M_q^2} \nonumber \\
 & & \quad \times \,\Big( \;\overline{\ell^c_+}\, (z) \;  \gamma^\mu \, \frac{1}{2}\,(1-\gamma_5) \left( \begin{array}{cc} 0_3 & 0_3 \\ 1_3 & 0_3 \end{array} \right) \ell_- (z) \;\Big[ \, \left( \begin{array}{cc} 0_3 & V_{{}_{CKM}} \\ 0_3 & 0_3 \\ \end{array}\right) \, \Big] \nonumber \\[2mm]
 & & \quad\;\;\,\, +  \,\overline{\ell^c_+}\, (z) \; \gamma^\mu \,\frac{1}{2}\,(1-\gamma_5) \left( \begin{array}{cc} 0_3 & 1_3 \\ 0_3 & 0_3 \end{array} \right) \ell_- (z) \; \Big[ \, \left( \begin{array}{cc} 0_3 & 0_3 \\ \overline{V}_{{}_{CKM}} & 0_3 \\ \end{array}\right)\,\Big] \nonumber \\[2mm]
 & & \quad\;\;\,\, + \, \overline{\ell^c_+}\,(x) \; \gamma^\mu \, \frac{1}{2}\, \Big( T_3 \, (1-\gamma_5)  - 2\;Q_\ell\, \sin^2 \theta_W \Big) \, \ell_-(z) \;\,  \Big[ \,2\, T_3 \,\Big] \; \Big) \, \Big] + \ldots \; .
\end{eqnarray}
Inspection yields for the decay constant $f_{\eta_{q_1\bar{q}_2}}$ of a pseudoscalar meson $\eta_{q_1\bar{q}_2}$
\begin{equation} i \, f_{\eta_{q_1\bar{q}_2}}  \longleftrightarrow \;  4 \,  N_c \, |g| \int \frac{d^4p}{(2\pi)^4} \;  \frac{(m_{q_1} + m_{\bar{q}_2})/2}{(p^2- m_{q_1}^2)(p^2- m_{\bar{q}_2}^2)} \; , \label{pseudconsteq1}
\end{equation}
being in accordance with the log.-divergent gap-equation Eq.\ (\ref{logdiveq1}) promoted by M.D.\ Scadron\footnote{The log.-divergent gap-equation should be understood here as a prescription to renormalize the original unrenormalized Lagrangean in replacing originally divergent quantitites by finite experimental numbers through a suitable choice of counter terms implying Eq.\ (\ref{logdiveq1}).
It is interesting to note that the previous result yields the extremly important sum-rule (resulting from the properties of the underlying integral) $(m_{q_1} - m_{\bar{q}_3}) \; f_{\eta_{q_1 \bar{q}_3}}  =  (m_{q_1} - m_{\bar{q}_2}) \; f_{\eta_{q_1 \bar{q}_2}} + (m_{q_2} - m_{\bar{q}_3}) \; f_{\eta_{q_2\bar{q}_3}}$
yielding e.g. $(m_u - m_s) \; f_{K^+} \; = \; (m_u - m_d) \; f_{\pi^+} + (m_d - m_s) \; f_{K^0}$.}.
As we will need it in the following we have now to dynamically generate the effective action describing the coupling of a scalar and two pseudoscalar mesons. The result is listed in the footnote \footnote{In the considered local limit we obtain:
\begin{eqnarray} \lefteqn{\frac{i}{1!} \, {\cal S}_{e\!f\!f} = \frac{i^3}{3!} \, \frac{3!}{1!\,2!} \, \left<0\right|\! T [\,{\cal S}^{\,Sq\bar{q}}_{int}\,{\cal S}^{\,Pq\bar{q}}_{int}\,{\cal S}^{\,Pq\bar{q}}_{int}\,]\left|0\right>_c\Big|_{\mbox{except $SPP\;$}} \; =} \nonumber \\
& \stackrel{!}{=} &
\int d^4z \;\; \sqrt{2} \;\;  g^2 \;\,e^{i\,\alpha} \,\; s_s \,  \; (- 4\; i \; N_c \; |g|) \; \int \frac{d^4p}{(2\pi)^4} \nonumber \\
 & \times & \Big\{ \mbox{tr}_F \Big[ \, S(z) \; \frac{1}{(p^2-M^2_q)}\; \{\, P^2(z)\,,\,M_q\, \} \; \frac{1}{(p^2-M^2_q)} \; \Big] \nonumber \\
 & & + \mbox{tr}_F \Big[ \, [\,S(z) \, , \, P(z)\,] \; \frac{1}{(p^2-M^2_q)}\; [\, P(z)\,,\,M_q\, ] \; \frac{1}{(p^2-M^2_q)} \; \Big] \nonumber \\
 & &  - \,\mbox{tr}_F \Big[ \,  \{ \, S(z)\, , \, M_q \, \} \; \frac{1}{(p^2- M_q^2)} [\, P(z)\,,\,M_q\, ] \; \frac{1}{(p^2-M^2_q)}\; [\, P(z)\,,\,M_q\, ] \; \frac{1}{(p^2-M^2_q)} \; \Big] \Big\}  +  \ldots \nonumber \\ \label{sppeq1}
\end{eqnarray}
Recalling our ``defining'' equation for pseudoscalar decay constants Eq.\ (\ref{pseudconsteq1}) the first two terms on the right-hand side of Eq.\ (\ref{sppeq1}) are equivalent to a $SPP$-interaction term, which one would obtain from a ``shifted'' quartic interaction Lagrangean with quartic coupling $\lambda$. The ``shifted'' Lagrangean is ${\cal L}(x) = - \frac{\lambda}{2} \,\mbox{tr}_F [  ((S(x) + i\,P(x) - D)(S(x) - i\,P(x) - D))^2 ] = \lambda \;\mbox{tr}_F [ \, (S(x) + i\,P(x))(S(x) - i\,P(x))(\{ S(x) , D\}+ i \,[ P(x) , D] ) ]+\ldots$. The quantity $D$ is the matrix (identified with decay constants of neutral pseudoscalar mesons) leading to spontaneous symmetry breaking according to the shift $S(x) \rightarrow S(x) - D$ and inducing quark-masses according to the relation $M_q = \sqrt{2}\; g \; s_s \; D$. The last term on the right-hand side of Eq.\ (\ref{sppeq1}) involving only commutators $[\, P(z)\,,\,M_q\, ]$ is proportional to the square of quark-mass differences and therefore small in the sense of the nonrenormalization theorem by M.\ Ademollo and R.~Gatto \cite{Ademollo:1964sr}.}.
In order to arrive at our final conclusions we can use the previous result to study the experimentally measured transition formfactors $f^{K^+\pi^0}_\pm(0)$ characterizing the process  $K^+\rightarrow \pi^0\,e^+\nu_e$ at zero four-momentum transfer. First we dynamically generate the respective effective action in the local limit displaying here only the for us relevant terms representing $W$-emission graphs and an exchange of a scalar $\kappa^+$-meson due to Partial Conservation of Vector Currents (PCVC)\cite{Schechter:1993tc} \footnote{The exchange of a vector meson $K^\ast$ is here disregarded, as it can contribute only marginally to the transition formfactor $f^{K^+\pi^0}_+(0)$ at zero four-momentum transfer, i.e. at most of the order of the nonrenormalization theorem by M.\ Ademollo \& R.~Gatto \cite{Ademollo:1964sr}, as the charge of the $K^+$ is solely generated due to photon-quark interactions.}:
\begin{eqnarray} {\cal S}_{e\!f\!f} & = & \int d^4z \;\; (-i\,e^{2\, i\,\alpha}) \; \left( - \frac{G_F}{\sqrt{2}} \right) \; \overline{V}_{us}\;\; \overline{e^{\,c}_+}\, (z) \; \gamma_\mu \, (1-\gamma_5) \; \nu_{e\,-}(z) \nonumber \\
 & \times &  \frac{1}{\sqrt{2}} \;\Bigg\{ \, \pi^0(z) \, \left( \, \frac{2\,|\,g|\,f_{K^+}}{m_u + m_s} \;\, \partial^\mu \, K^+(z)\right) \, - \, K^+(z) \, \left( \, \frac{2\,|\,g|\,f_{\eta_{u\bar{u}}}}{m_u + m_u} \;\, \partial^\mu \,\pi^0(z)\right) \nonumber \\
 & & + 4 i \, N_c \, |\,g|^2 \, (m_s - m_u)^2 \; K^+(z) \, ( \partial^\mu \, \pi^0(z)) \int \frac{d^4p}{(2\,\pi)^4}\, \frac{1}{(p^2-m^2_s)(p^2-m^2_u)^2} \nonumber \\
 & & + \frac{\lambda}{g^2} \, m_s \, \frac{(m_s - m_u)}{m^2_{\kappa^+}} \, \frac{2\,|\,g|\,f_{K^+}}{m_u + m_s} \, \Big( \pi^0(z) \, (\partial^\mu \, K^+(z))  +  K^+(z) \, (\partial^\mu \, \pi^0(z) )   \Big)   \Bigg\} \nonumber \\
 & + & \mbox{$K^\ast$-exchange} + \ldots \, .
\end{eqnarray}
From this result we can read off the desired transition formfactors $f^{K^+\pi^0}_\pm(0)$ at zero four momentum transfer. Displaying only terms being of relevant order in the scale  $\delta = (m_s/m_u) - 1\simeq 0.44\;$  according to the nonrenormalization theorem of M.\ Ademollo and R.~Gatto \cite{Ademollo:1964sr} we obtain $f^{\,K^+\pi^0}_+(0)= 1 + O(\delta^2)$ and
\begin{eqnarray} \lefteqn{f^{\,K^+\pi^0}_-(0) - O(\delta^2) = \frac{\lambda}{g^2} \, m_s \,  \frac{(m_s - m_u)}{m^2_{\kappa^+}} \, \frac{2\,|\,g|\,f_{K^+}}{m_u + m_s} =}\nonumber \\
 & = & e^{2 i \alpha} \;  \frac{\lambda}{g^2} \; \frac{2\,\delta \,(1+\delta)}{(2+\delta)}  \frac{|m_u||f_{K^+}|}{m^2_{\kappa^+}}\;|g|  \stackrel{!}{=} e^{2 i \alpha} \, \frac{2\,\delta \,(1+\delta)}{(2+\delta)}  \frac{|m_u||f_{K^+}|}{m^2_{\kappa^+}}\;\frac{4\pi}{\sqrt{3}}  \; . \label{lasteq1}
\end{eqnarray}
On the right-hand side of this equation we used that M.D.\ Scadron's log.-div.\ gap-equation Eq.\ (\ref{logdiveq1}) in combination with Eq.\ (\ref{pseudconsteq1}) implies $|g|= 2\pi/\sqrt{N_c}=2\pi/\sqrt{3}$, and that there holds $\lambda \simeq 2 \, g^2$ according to a one-loop dynamical generation \cite{Delbourgo:1993dk}.
In using the experimental values \cite{Eidelman:2004wy} $|f_{K^+}|\simeq 159\;\mbox{MeV}/\sqrt{2}$ and $m_{\kappa^+} \simeq 797$ MeV we produce with the help of the last line of Eq.\ (\ref{lasteq1}) the following {\bf Table 1}:\\[1mm]
\mbox{} \hfill \begin{tabular}{|c||c|c|c|c|c|c|c|c|c|} \hline
$f^{\,K^+\pi^0}_-(0)/e^{2 i \alpha}$ & 0.050 & 0.102 & 0.125 & 0.148 & 
0.200 & 0.225 \\ \hline
$\delta$ for $|m_u|=337$ MeV       & 0.1098 & 0.2149 & 0.2591 & 0.3023 &
0.3965  & 0.4404 \\
$\delta$ for $|m_u|=\;\;\;3$ MeV    & 7.274 & 14.11 & 17.12  & 20.12 & 
26.89 & 30.14\\ \hline
\end{tabular} \hfill \mbox{}\\[2mm]
\noindent Inspection of the constituent quark mass case $|m_u|\simeq 337$ MeV reveils that the experimentally measured {\em negative} transition formfactor ratio $f^{\,K^+\pi^0}_-(0)/f^{\,K^+\pi^0}_+(0) \simeq - 0.125 \pm 0.023$ \cite{Eidelman:2004wy} can be only accomodated for $e^{2 i \alpha}<0$, while for reasonable values of $\delta$ and $m_\kappa$ experiment seems to suggest the extreme PT-symmetric \cite{Bender:1998ke} case $\alpha\simeq -\pi/2 + 0$ yielding an imaginary PT-symmetric Yukawa-coupling $g=-i\,2\pi/\sqrt{3}$ and a Symanzik-like quartic coupling $\lambda\simeq 2\,g^2 = - 8\pi^2/3<0$, as obtained already earlier by the author, when ``deriving'' the Lagrangean of the Quark-Level Linear Sigma Model from the Lagrangean of QCD \cite{Kleefeld:2002au} (See also Ref. \cite{Kleefeld:2004jb}). Finally it is interesting to consider our rough estimate for the experimentally yet badly determined mass of the $\kappa(800)$ scalar resonance (biased by $K^\ast_0(1430)$) as a function of $\delta$. For $|m_u|\simeq 337$ MeV and $f^{\,K^+\pi^0}_-(0)= - 0.125$ we obtain the following {\bf Table 2}: \\[1mm]
\mbox{} \hfill \begin{tabular}{|c||c|c|c|c|c|c|c|} \hline
$\delta$        & 0.10  & 0.20 & 0.26 & 0.30 & 0.40 & 0.44 & 0.50  \\ 
\hline
$m_\kappa$ [MeV] & 480.0 & 692.7  & 798.5 & 863.6 & 1013.1 & 1068.69 &
1148.7 \\\hline
\end{tabular} \hfill\mbox{}\\[2mm]
\noindent Hence, semileptonic decays of pseudoscalar mesons can not only be used to reveil
the seemingly non-Hermitian nature of a theory of strong interaction with a sizable amount of scalar
confinement, they also may be used to ``measure'' badly known experimental quantities like the masses of light and heavy scalar resonances.\\[1mm]
{\small \indent This work is dedicated to M.D.\ Scadron and G.\ Rupp. It has been supported by the
{\em Funda\c{c}\~{a}o para a Ci\^{e}ncia e a Tecnologia} \/(FCT) of the {\em Minist\'{e}rio da Ci\^{e}ncia e da Tecnologia (e do Ensino Superior)} \/of Portugal, under Grants no.\ PRAXIS XXI/BPD/20186/99, SFRH/BDP/9480/2002, POCTI/\-FNU/\-49555/\-2002, and POCTI/FP/FNU/50328/2003.}

\end {document}